
\input harvmac.tex
\noblackbox
\lref\LM{A. Leclair and G. Mussardo, ``Finite temperature correlation
functions in integrable QFT'', hep-th/9902075.}
\lref\Zamo{Al. Zamolodchikov, Nucl. Phys. B342 (1990) 695.}
\lref\LS{F. Lesage and H. Saleur, Nucl .Phys. B490 (1997) 543.}
\lref\LLSS{A. LeClair, F. Lesage, S. Sachdev, and H. Saleur,
Nucl. Phys. B482 (1996) 579.}
\lref\FSW{P. Fendley, H. Saleur and N. P. Warner, Nucl. Phys. B430 (1994) 577.}

\Title{USC-99-05}
{\vbox{
\centerline{A comment on Finite Temperature }
\vskip1cm
\centerline{Correlations in Integrable QFT }}}

\centerline{ H. Saleur}

\bigskip\centerline{Department of Physics}
\centerline{University of Southern California}
\centerline{Los Angeles, CA 90089-0484}

\vskip .3in
I discuss and extend the recent proposal of Leclair and Mussardo for 
finite temperature correlation functions in integrable QFTs. I give 
further justification for its validity in the case of one point functions 
of conserved quantities. I also argue that the proposal is not correct for
two (and higher) point functions, and give some counterexamples to 
justify that claim.

\Date{09/99}

In a recent paper \LM, A. Leclair and G. Mussardo have come up with 
a very interesting suggestion to compute correlation functions in integrable
quantum field theories at finite temperature. In a nutshell, what  they
propose is to identify the usual particle and hole excitations over the
``thermal
ground state'' \LLSS, with their non trivial dressed energies and momenta
(as obtained by solving the thermodynamic Bethe ansatz equations), but
still take, for the physical operators, the undressed form factors. 

The proposal in \LM\ is quite exciting, as finite temperature
correlation 
functions are quantities of the highest interest (in particular for
comparison with experiments). However, it is fair to say that no
real demonstration is provided in \LM\ except for some one
point functions, and  that the formula in \LM\ are  merely a
tentalizing  guess. 

I will 
argue in this note that, while the formula proposed in \LM\ for one point
functions are, at least for some operators (the densities of conserved quantities),  correct, the ones for two and higher point functions 
generally do not hold. The reason for this is 
quite simple and physical: the formula proposed in \LM\ does not take
into
account the effects of the dressing in the form-factors. This is fine 
for one point functions: since the {\sl same} multiparticle states 
are on the left and on the right side of the correlator, there is no
motion of the thermal ground state \LLSS\  due
to interactions, and the formula for bare form factors can still be used. In
contrast, for two point functions, one has to consider intermediate
states where particle or holes have been created; this leads to a 
displacement of the thermal ground state, and a dressing of the form-
factors, that has to be taken into account - there is  just no reason for
the bare form factors to still be relevant there.

One of the difficulties in assessing the validity of \LM\ is that very few 
finite temperature correlators are known exactly in integrable theories,
and that the expressions proposed in \LM, even if conceptually quite simple,
are nevertheless very hard to compute explicitely \foot{A possibility might be 
to investigate the low temperature limit, and compare it 
with results obtained in \ref\Subir{S. Sachdev and K. Damle, Phys. Rev. Lett. 78 (1997) 943.}
, but I won't do 
this here.}. However, as I will argue below, the argument presented in \LM\ is quite general, and holds just as well for correlators evaluated in other thermodynamic ensembles. The case of a chemical potential at $T=0$ turns out to be particularly simple, and will allow me to put the two point functions of \LM\ to a serious test, that they unfortunately fail.

Let me now proceed and discuss one point functions first. 
I will only consider 
 operators $O(x)$ (I refer to them, a bit incorrectly, 
as conserved quantities) for which the quantity $O=\int O(x)dx$ acts diagonally on multiparticle states, with one particle
eigenvalues
$o(\theta)$. Examples of this include the energy and momentum
operators,
and after a very slight modification, the trace of the stress energy
tensor considered in \LM.  By translational invariance, it thus follows
that
$$
{\langle
\theta_n,\ldots,\theta_1|O(x)|\theta_1,\ldots,\theta_n\rangle\over
\langle
\theta_n,\ldots,\theta_1|\theta_1,\ldots,\theta_n\rangle}={1\over L}\sum_{i=1}^n
o(\theta_i),
$$
$L$ the system length. 

I want now to compute the  average of $O$ at finite temperature,
and in the presence of possible other thermodynamic couplings. 
I will restrict for notational 
simplicity to  a
theory with a single particle, and consider, in addition to the
temperature, the presence of a chemical potential. Generalizations
are quite straightforward.
Due to the non trivial S matrix of the theory,
the density of allowed states
 $P$ ($P=\rho+\rho^h$) obeys (I set $\hbar=1$ here):
\eqn\tbai{2\pi P(\theta)=Lm\cosh\theta+ 2\pi \Phi\star \rho(\theta),}
where I defined $f\star g(\theta)=\int
  f(\theta-\theta')g(\theta') {d\theta'\over 2\pi}$. In \tbai,
the kernel 
$\Phi={1\over 2i\pi}{d\over d\theta}\ln S$, where 
$S$ is the scattering matrix.

The average of $O$ is computed using the thermodynamic Bethe ansatz
method:  configurations are weighed with a Boltzmann weight
$\exp\left[
-{E-TS-\mu N\over T}\right]$ where $T$ is the temperature, $S$ the
entropy,
$\mu$ the chemical potential, and $N$ the number of particles, and
saddle point equations are written, leading to the TBA equations. 
Introducing the pseudo energy $\epsilon$ defined by
${\rho^h\over\rho}=e^\epsilon$,  the TBA equation at temperature $T$ (
$T={1\over R}$) read
\eqn\tbaii{\epsilon(\theta)=m R\cosh\theta
  -\Phi\star\ln\left(1+e^{-\epsilon+\mu R}\right),}
It can  be shown with the usual argument \Zamo\ that $2\pi
P=L\left.{\partial\epsilon\over\partial R}\right|_{\mu R~fixed}$. From this the average of $O$
follows:
\eqn\tbaiii{\eqalign{
\langle O(x)\rangle=&{1\over L}\int o(\theta)\rho(\theta)d\theta\cr
=&\int o(\theta)f_-(\theta)\left.{\partial \epsilon\over \partial
R}\right|_{\mu R~fixed}{d\theta\over 2\pi},\cr}}
with $f_-={1\over 1+e^{\epsilon-\mu R}}$. I can now solve the TBA
equation
iteratively to obtain the expansion
\eqn\tbaiv{\eqalign{ \langle O(x)\rangle=\int o(\theta)f_-(\theta)m\cosh\theta
  {d\theta\over 2\pi}+\int o(\theta)f_-(\theta){d\theta\over 2\pi}
\int \Phi(\theta-\theta')f_-(\theta')m\cosh\theta' {d\theta'\over
  2\pi}\cr
+\int o(\theta)f_-(\theta){d\theta\over 2\pi}
\int \Phi(\theta-\theta')f_-(\theta'){d\theta'\over 2\pi}
\int \Phi(\theta'-\theta'')
f_-(\theta'')m\cosh\theta'' {d\theta''\over
  2\pi}+\ldots\cr}}
Notice that this expression in terms of $o,\Phi$ and the 
filling fraction $f_-$ is completely general. It would hold for 
a system with more thermodynamic couplings, provided these couplings
involve conserved quantities. The expression \tbaiv\ also generalizes
 easily to cases with several types of particles.

 I will now \tbaiv\ compare with the formula proposed in \LM:
\eqn\lmguess{\langle O(x)\rangle=\sum_{n=0}^\infty {1\over n!}
\int {d\theta_1\over 2\pi}\ldots{d\theta_n\over 2\pi} \prod_{i=1}^n
f_-(\theta_i)\langle
\theta_n\ldots\theta_1|O(x)\theta_1\ldots\theta_n\rangle_{conn},}
and argue that it is the same. Of course, in \LM, formula \lmguess\ is
written only in the case of finite temperature. However, all arguments 
presented there rely on cancellations between numerators and denominators,
 together with hints from the free case and the one point function 
of the stress energy tensor, and these are all 
features which generalize to more complicated thermodynamic averages, like the ones involving a chemical potential. In \lmguess, the connected form factor is obtained in principle 
by using the crossing formula for non coinciding arguments,
and then getting rid of all the diverging or ill defined terms 
as the arguments are sent to one another. 

The key point  here is that  the values of the connected form-factors can be understood 
generally as follows (this argument  is quite similar to
Balog's construction \ref\Balog{J. Balog, Nucl. Phys. B419 (1994)
480.}). To
start, I need to make some
remarks on the normalization of states. Consider the scalar product $\langle \theta|\theta\rangle$, which, formally,
is equal to $2\pi \delta(0)$, since the normalization used in the construction of the 
asymptotic states is $\langle \theta|\theta'\rangle=2\pi\delta(\theta-\theta')$.  Of course the
symbol $\delta(0)$ does not make much sense, and has to be regularized properly. The 
way to do this is to consider the completude relation, $1=\int {d\theta\over 2\pi} |\theta\rangle\langle\theta|$. 
This relation hides the cancellation of two terms, the meaning of which is 
easier to see by introducing a finite length $L$ in the system. First, the state $|\theta\rangle$ is actually not 
normalized, because of the $\delta(0)$ term mentioned just before; the integral should therefore 
involve instead the ket $|\theta\rangle\over\sqrt{\langle \theta|\theta\rangle}$ and similarly for the 
associated bra. Second, allowed rapidities for a particle are {\sl not} uniformly distributed; rather,
they have a  density given by $2\pi P=m L \cosh\theta$, and the  completude relation should involve therefore
an integral $\int P(\theta)d\theta$. For the completude relation to be equivalent to 
the one we used so far, which preserves the scalar product of states, we  thus need to have 
\ref\LMSS{A. Leclair, G. Mussardo, H. Saleur, and S. Skorik, Nucl. Phys. B453 (1995) 581.}
\eqn\deli{ \langle \theta|\theta\rangle 
=  mL\cosh\theta.}

The argument can be generalized to the case with several rapidities. For two particles, 
the Bethe equations read:
\eqn\bet{\eqalign{2\pi n_1= & mL\sinh\theta_1 +{1\over i}\ln S(\theta_1-\theta_2)\cr
2\pi n_2= & mL\sinh\theta_2 +{1\over i}\ln S(\theta_2-\theta_1).\cr}}
From this, we deduce that
\eqn\delii{\eqalign{\langle \theta_2,\theta_1|\theta_1,\theta_2\rangle=& \hbox{Det }\left|\eqalign{mL \cosh\theta_1 +\Phi(\theta_{12})&~~~~~~ -\Phi(\theta_{12})\cr
-\Phi(\theta_{12})~~~~~~& mL\cosh\theta_2+\Phi(\theta_{12})\cr}\right|\cr
= &  m^2 L^2\cosh\theta_1\cosh\theta_2+mL\Phi(\theta_{12})(\cosh\theta_1+\cosh\theta_2).\cr}}
The formula generalizes to an arbitrary number of rapidities. As could have been expected, it coincides with 
the well known formula for the norm of Bethe states derived by Gaudin and by Korepin in the context of the 
XXZ and other models (see \ref\Korbook{V.E. Korepin, N.M. Bogoliubov and
A.G.Izergin, ``Quantum inverse scattering method and correlation functions'', Cambridge University Press, 1993.}).

I can  now get back to \tbaiv\ and \lmguess. It is clear that the
first term of the expansions match, due to \deli\ 
\eqn\ji{\langle \theta|O(x)|\theta\rangle\equiv\langle
    \theta|O(x)|\theta\rangle_{conn}= o(\theta) m\cosh\theta .}

For the two particle form factor, let me  define:
\eqn\jii{\eqalign{\langle
  \theta_2,\theta_1|O(x)|\theta_1,\theta_2\rangle_{conn}=&
\langle
  \theta_2,\theta_1|O(x)|\theta_1,\theta_2\rangle -\langle
  \theta_2|O(x)|\theta_2\rangle_{conn}\langle\theta_1|\theta_1\rangle\cr
&-\langle
  \theta_1|O(x)|\theta_1\rangle_{conn}\langle\theta_2|\theta_2\rangle.\cr}}
Here, the scalar products (eg $\langle \theta_1|\theta_1\rangle$) have to
be computed in the presence of the other particle ($\theta_2$), something I left implicit in the notations for simplicity. From the
general arguments 
explained previously,  $\langle \theta_1|\theta_1\rangle=
mL\cosh\theta_1+
\phi(\theta_{12})$. It thus follows that
\eqn\jiii{\langle
  \theta_2,\theta_1|O(x)|\theta_1,\theta_2\rangle_{conn}=
m\Phi(\theta_{12})\left[ \cosh\theta_1 o(\theta_2)+\cosh\theta_2
  o(\theta_1)\right],}
again ensuring a matching between \tbaiv\ and \lmguess. A 
general formula for  connected form-factors follows easily by 
extending  \jii\
from $n=2$ to arbitrary values of $n$: simply subtract from the initial
form factor all the possible contractions, being careful to evaluate 
scalar products of states in the presence of all other particles -
they are all expressed as   various minors of the same  initial determinant. The net
result is simply:
\eqn\jiv{\langle
  \theta_n,\ldots,\theta_1|O(x)|\theta_1,\ldots,\theta_n\rangle_{conn}=
m\Phi(\theta_{12})\Phi(\theta_{23})\ldots\Phi(\theta_{n-1,n})
  \cosh\theta_1
o(\theta_n)+\hbox{ permutations},}
and clearly ensures the coincidence of \tbaiv\ and \lmguess.

I am of course not pretending that 
this argument is a proof of \lmguess, but I believe it could 
be sufficiently strenghtened, maybe along the lines of \Balog. 
As it stands, it certainly gives further support to \lmguess, and to my claim
that the guess of \LM\ should be considered for more general thermodynamic ensembles.

I shall now argue that, unfortunately,
the formula proposed in \LM\ for the two
particle
correlator is probably not correct. Rather than give general arguments,
I will   present a simple counterexample, in the case where $T=0$ but there is 
a non trivial chemical potential.
  
Consider indeed the free boson with
hamiltonian
\eqn\hamil{H={1\over 2}\int  \left[ 8\pi g\Pi^2+{1\over 8\pi g}
    \left(\partial_x\phi\right)^2\right]dx.}
I will add to this hamiltonian a term ${V\over 4\pi} \int 
\partial_x\varphi dx$. Properties in the presence of this coupling are 
straightforward to evaluate by a simple shift of the bosonic
field. Introducing chiral components $\varphi,\bar{\varphi}$, one finds ($\partial_x\phi=\partial_z\varphi+
\partial_{\bar{z}}\bar{\varphi}$)
\eqn\aver{\eqalign{\langle\partial_z\varphi\rangle=& Vg\cr
\langle\partial_{z_1}\varphi\partial_{z_2}\varphi\rangle=& (Vg)^2
-{2g\over (z_1-z_2)^2},\cr}}
where I have set $z=x+iy$, $y$ the imaginary time (called $t$ in \LM). 

On the other hand, we can consider the free boson as the UV limit
of the sine-Gordon theory, where the hamiltonian \hamil\ is
supplemented
by an interaction term $\lambda\int  \cos\phi dx$. 
Provided we consider physics at a scale much smaller than the 
correlation length induced by this perturbation, ${1\over M}\propto
\lambda^{-2+2g}$, we will observe results similar to 
the free boson case. Equivalently, we  expect to be able describe the
properties
of the free boson theory using a massless scattering description
\FSW, with
purely right and left moving particles obtained by taking the large
rapidity limit of the usual solitons, antisolitons and breathers of
the theory. The parametrization of the energy I will use in that
limit is $e=\pm p= e^\theta$, where I have set an arbitrary mass scale equal to unity.
The correlators \aver\ should therefore be obtainable
using the formulas proposed in \LM. I will show that, actually,
only the one point function is obtained.

For simplicity, I restrict to
the case $T=0$. The field $V$ then leads to the creation 
of a pair of  Fermi seas  of left and right moving massless solitons: in the following, I will concentrate 
on the right moving sector only.  The average of
$\partial_z\varphi$ is directly related with the number of 
solitons in the ground state 
$$
{1\over 2\pi} \langle \int  \partial_z\varphi dx\rangle =\langle N\rangle
$$
This quantity fits exactly in the context described previously. The
two particle kernel is given by the soliton soliton S matrix, and,
since we are at temperature $T=0$, the function $f_-$ becomes very
simple: $f_-=1$ for particles in the sea, $f_-=0$ otherwise. If we
consider the operator $\partial_z\varphi$, it couples only to
right moving particles, for which the sea is $\theta\in (-\infty, A)$,
$A$ a Fermi rapidity. The average of $\partial_z\varphi$  follows from 
the appropriate generalization of \tbaiv:
\eqn\avi{\langle \partial_z\varphi\rangle=\int_{-\infty}^A 2\pi e^\theta{d\theta\over 2\pi}+
\int_{-\infty}^A 2\pi {d\theta\over 2\pi}\int_{-\infty}^A
\Phi(\theta-\theta')
e^{\theta'}{d\theta'\over 2\pi}+\ldots.}

Expressions for the form-factors of the operator $\partial_z\varphi$ 
are well known: we can, in principle, directly 
compute the connected ones, 
and show agreement with the general formula \jiv. For instance, 
the two particle form-factor gives
\eqn\checki{\langle \theta|\partial_z\varphi|\theta\rangle=2\pi  e^\theta,}
in agreement with (the massless limit of) \ji\ for $o(\theta)=2\pi$, the value of the integral of $\partial_z\varphi$ 
for a soliton. I have checked similarly the formula for the connected 
four particle form-factor \jiii\ for simple values of the coupling $g$.

I now turn to the two point function of the operator
$\partial_z\varphi$, 
and show that the expression proposed for it in \LM\ this time does not work. 
Recall the proposal of \LM:
\eqn\lmi{\eqalign{\langle O(x,y) O(0,0)\rangle_R &=
\left(\langle O\rangle_R\right)^2 +\sum_{N=1}^\infty {1\over N!}
\sum_{\sigma_i=\pm 1} \int {d\theta_1\over 2\pi}\ldots{d\theta_N\over
  2\pi}
\prod_{j=1}^N f_{\sigma_j}(\theta_j)\cr
&  \exp\left(i\sigma_j
    (x\tilde{k}_j+ iy\tilde{e}_j)\right)
\left|\langle 0|O(0)|\theta_1,\ldots,\theta_N
\rangle_{\sigma_1,\ldots,\sigma_N}\right|^2.\cr}}
Here, $f_+={1\over 1+e^{-\epsilon}}$, $\tilde{e}=R\epsilon$ is the 
dressed excitation energy, while $\tilde{p}$ is the dressed excitation
momentum. Finally,
$$
\langle
0|O(0)|\theta_1,\ldots,\theta_N\rangle_{\sigma_1,\ldots,\sigma_N}\equiv
\langle
0|O(0)|\theta_1-i\pi(\sigma_1-1)/2,\ldots,\theta_N
-i\pi(\sigma_N-1)/2\rangle
$$
The physical meaning of \lmi\ is more transparent than
the formula. Correlations at temperature $T$ should be determined
by processes involving excitations over the thermal ground state,
with dressed energy and momentum \LLSS. These excitations can be of two
types, particles or holes (the variable $\sigma$ in \lmi). A sum as \lmi\ is thus expected; the key
proposal
of \LM\ is that the form factors for the physical excitations are the 
same as the bare ones. 

In the case we consider, the situation simplifies considerably. 
First, let me recall the structure of the excitations \LS. It is 
conveniently represented in the following table:
\eqn\tabl{\eqalign{ \hbox{holes }&:~ \epsilon_+^h(\theta),~ \theta\leq
    A\cr
\hbox{solitons }&:~\epsilon_+(\theta),~\theta\geq A\cr
\hbox{antisolitons }&:~\epsilon_-(\theta)\geq (1-g) V,~\theta\hbox{ arbitrary}\cr
\hbox{breathers }&:~\epsilon_n(\theta)\geq ngV,~\theta\hbox{
    arbitrary}\cr}}
The excitation energies $\epsilon$ have non trivial expansions given
in \LS. 

The physical processes involved in the two point function of
$\partial_z\varphi$ are of three basic types: creation of a pair
particle hole (ie take a soliton in the Fermi sea and move it
outside);
creation of a pair soliton (above the Fermi sea) antisoliton,
and creation of a breather. The latter two processes have thresholds.
Since the quantities $\tilde{e}$ and $\tilde{p}$ 
have no singularity, and the bare form-factors do not know anything
about these thresholds, it is immediately quite obvious that the
proposal
of \LM\ cannot be true: it would lead to singularities in 
the Fourier spectrum of the correlator, in sharp contrast
with the result expected from \aver. 

Nevertheless, it is probably reasonable to make this counter example
more explicit. 
First, let me show that the formula is in fact right for the free case
$g={1\over 2}$. In that case, things simplify for several reasons:
there are no breathers, the excitation energies have simple
expressions $\epsilon_+^h={V\over 2}-e^\theta$,
$\epsilon_+=e^\theta-{V\over 2}$, $\epsilon_-=e^\theta+{V\over 2}$, 
and only the two particle form factor of $\partial_z\varphi$ is non
zero, $\langle 0|\partial_z\varphi|\theta_1\theta_2\rangle=2i\pi
e^{\theta_1/2}
e^{\theta_2/2}$. The connected term in expression \lmi\ reads then
$$
\eqalign{ &\int_{-\infty}^A {d\theta\over 2\pi}\int_A^\infty {d\theta_2\over 2\pi}
\left| 2\pi e^{\theta_1/2}e^{\theta_2/2}\right|^2
e^{iz(e^{\theta_2}-e^{\theta_1})}\cr
&+ \int_{-\infty}^\infty {d\theta_1\over 2\pi}\int_A^\infty {d\theta_2\over 2\pi} \left|
2i\pi e^{\theta_1/2}e^{\theta_2/2}\right|^2
e^{iz(e^{\theta_1}+e^{\theta_2})}\cr
&= -{1\over z^2}\left[1-\exp(e^A z)\right]-{1\over z^2} \exp(e^A
z)=-{1\over z^2}\cr}
$$
the expected result \aver\ for $g={1\over 2}$. 

In the general case, I will consider the Fourier transform 
of the two point function, $S(\omega)=\int {dx\over 2\pi} e^{-i\omega x} \langle \partial_z\varphi(x,y)\partial_z\varphi(0)\rangle_{y\to 0}=
2g \omega$ ($\omega>0$). I will 
also restrict to the case  of low frequencies (that is, $\omega$ smaller than any of the thresholds), where only the 
particle hole processes contribute; if formula \lmi\ fails in that 
case, it will be enough to show it is not correct. 

Below the lowest thresholds, $S(\omega)$, according to formula \lmi, is given by a sum of terms 
$S_n$ corresponding to  processes 
with $n$ particles and  $n$ holes. When $V=0$ and we know formula \lmi\ is correct, each of the $S_n$'s
is linear in $\omega$, $S_n=c_n\omega$. In that case, there are no thresholds, so the $S_n$ have to 
be supplemented by the terms corresponding to the other processes, each of which is also proportional to $\omega$. The infinite sum of all these $\omega$ terms reproduces the desired behaviour $S(\omega)=2g\omega$;
as checked in \ref\LSS{F. Lesage, H. Saleur, and S. Skorik, Nucl. Phys. B474 (1996) 602.} the convergence of this sum is in fact very quick for $g$  not too close to 1. When $V\neq 0$, formula \lmi\ predicts correctly
(this simply follows from dimensional analysis) that all the $S_n$'s now have the form $S_n=\omega f_n\left({\omega\over V}\right)$, with  $c_n=f_n(\infty)$. Let us now consider the limit
of small frequencies at finite $V$: in that case, we are instead  exploring the behaviour of the 
functions $f_n$ in the  limit of very small argument.  Since the
contribution
at frequency $\omega$ is determined by rapidities $\theta_i>A$ of
holes and $\theta'_i<A$ of particles such that
\eqn\jvii{\omega=\sum_{i=1}^n
  \epsilon_+^h(\theta_i)+\sum_{i=1}^n\epsilon_+(\theta_i'),}
clearly, for small $\omega$, the particles and holes have to be
closer
and closer to the Fermi rapidity. The two particle contribution
for instance, which reads in general
$$
\int_{-\infty}^A {d\theta_1\over 2\pi}\int_A^\infty {d\theta'_1\over 2\pi} \left|
\langle \theta'_1|\partial_z\varphi|\theta_1\rangle\right|^2
\delta\left[\omega-\epsilon_+^h(\theta_1)-\epsilon_+(\theta'_1)\right]  
$$
becomes in the limit of small $\omega$, $ {1\over (2\pi)^2}|\langle
A\partial_z\varphi|A\rangle|^2
{\omega\over|\dot{\epsilon}_+(A)\dot{\epsilon}
_+^h(A)|}$ (dots denote derivatives with respect to the rapidity 
variable). In fact, for the limit $\lim_{\omega\to 0}
{S(\omega)\over\omega}$, this two particle contribution is the {\sl
  only} one to  consider. There are two reasons for that: one is that
the 
next contribution is down by an $\omega^2$ term due to phase space
considerations; the other is that in the limit $\omega\to 0$, the
rapidities 
of the $n$ particles and $n$ holes are all compressed towards $A$, and
form-factors like eg $\langle A,A|\partial_z\varphi|A,A\rangle$ vanish
due to 
the general behaviour under rapidity exchanges ($S(0)=-1$). It follows that,
according to \LM, one would find
\eqn\necess{\lim_{\omega\to 0}{S(\omega)\over|\omega|}=^?= {1\over (2\pi)^2}{|\langle
    A|\partial\varphi|A\rangle|^2\over |\dot{\epsilon}_+(A)|
    |\dot{\epsilon}_+^h(A)|}.}
The right hand side can be evaluated using the detailed results of \LS\ for the filling of the ground 
state etc. One finds (for dimensional reasons, this has to be a pure, $V$ independent number)
\eqn\rhs{\hbox{RHS }= {1\over 2\pi} {g^{1-2g\over 1-g}\over 1-g} \left({\Gamma[g/2(1-g)]\over
    \Gamma[1/2(1-g)]}\right)^2.}
Except when $g={1\over 2}$, the right hand side is off the exact
result (it should be equal to $2g$) by a finite amount, demonstrating that \lmi\ is not correct in that case.

The reader might worry here that the whole argument assumes somehow absolute convergence of the series in \lmi, and that maybe one would get the correct limit in \necess\ by first summing over all contributions, 
then letting $\omega\to 0$. The answer to this is that I have considered the limit \necess\ 
only to make things as clear as possible. We are of course interested not only in this limit, but in the behaviour of $S(\omega)$ on the whole real axis, where we have to recover $S(\omega)=2g\omega$ for \lmi\ to be correct. As argued above, only the one particle hole contribution has a term linear in $\omega$: other terms start with higher powers in $\omega$, and it is easy to see that only a finite number of them contribute
to a given order in $\omega$.  The series representing $S(\omega)$ according to \lmi\ is thus of the form
$\sum d_k \omega^k$, with all the $d_k$'s finite (eqn. \rhs\ means that $d_1$ is not the right one; one can also check that $d_3$ is not the right one - that is, it does not vanish - etc). For my argument to be spoilt by a convergence problem, one would need this series to diverge and to somewhat ``represent'' the simple 
linear term $2g\omega$; besides the fact that I have found no indication of divergence numerically (that is,
convergence seems as good as in \LSS), I do not think such a scenario is likely at all.

The reader might also be surprised by the fact that \lmi\ is right for $V=0$ but not $V\neq 0$. The point is once again that \lmi\ leads to a representation of $S(\omega)$ as a sum of terms of the form $\omega f_n\left({\omega\over V}\right)$; validity in the  case $V=0$ is a statement about the sum of $f_n(\infty)$'s, while invalidity in the case $V\neq 0$ is a statement about the general shape of the sum of $f_n$'s at finite argument.  
 
The  physical origin of the failure of \lmi\ is easy to trace back to the dressing effects. In fact, in
\LS, another approach to compute correlators at $T=0$ with 
a field coupled to the $U(1)$ current was proposed. In this paper, it is recognized that dressed
excitations must have dressed form-factors; an expression was
 proposed in particular
for the low energy behaviour of the particle hole form factor, which
reproduced $S(\omega)$ correctly.

\bigskip

I should stress that I  have not found  simple counterexamples in the context originally
considered in \LM, of a theory with a temperature and no chemical potential. However, I believe the argument presented in \LM\ generalizes to any thermodynamic average, so this counterexample at $T=0$ with a chemical potential
is nevertheless a good, alas negative, test, of their result for two and higher point functions. Some easy things can be said about the $T\neq 0$ case
 however.
One of them has to do with {\sl thresholds}. Consider for instance
the massless limit of the sine-Gordon model once again, and its massless scattering description. At $T\neq 0$,
it is one of the striking features of the interactions that
the energy necessary to add  a particle has a gap: for instance, the energy 
to create a one-breather in the attractive regime is $T\epsilon_1\geq T\ln 3$; similarly the energy gained by destroying such a breather is larger or equal to this number (this follows simply from the solution of the TBA equations). As a result, if we consider again the correlator  of 
$\partial_z\varphi$, 
processes involving the one-breather have thresholds (recall that the 
bare form-factors of $\partial_z\varphi$ with an even number 
of one-breathers are zero). Since the form factors in the sum \lmi\ are the bare ones, they know 
nothing about these thresholds; as for the other pseudo energies, they 
have no singularity at the position of these thresholds. It follows 
that, according to \lmi, the Fourier transform of the two point 
function of $\partial_z\varphi$ would
 exhibit singularities at finite values of the frequency $\omega$. This is 
of course in contradiction with the simple form of this  two point function
, that follows from conformal invariance, and indicates once again that the proposal  in \LM\ is generally not correct.

In conclusion, I believe that the formulas of \LM\ for the two and 
higher point functions are in general incorrect because they don't take properly 
into account the dressing of the effective vacuum created by finite thermodynamic couplings. As for the one point function of conserved quantities, the formula of \LM\ 
looks very reasonable when compared with the result of the TBA, and presumably could be rigorously proven by a more serious analysis of connected form-factors than the one I have presented here. I am not sure about the one point functions
of non conserved quantities - the examples studied in \LM\ are quite convincing, but further work is probably needed to settle the issue.

\bigskip
\noindent{\bf Acknowledgments}: I thank A. Leclair, F. Lesage and G. Mussardo 
for many discussions on this problem. This work was supported by the DOE
and the NSF (under the NYI program). 

\listrefs

\bye